\documentstyle[12pt,cite,epsfig,psfrag]{article}

\newcommand {\Tr}{\mbox{Tr}}
\begin{document}
\baselineskip=0.6cm
\begin{titlepage}
\begin{center}
\hfill WITS-CTP-070\\
\vskip .2in
{\Large \bf Exact Large $R$-charge Correlators in ABJM Theory}
\vskip .6in
{\bf Tanay K. Dey\footnote{e-mail: Tanay.Dey@wits.ac.za}}\\
\vskip .1in
{National Institute for Theoretical Physics,\\
Department of Physics and Centre for Theoretical Physics,\\
University of the Witwatersrand,\\
Wits, 2050,\\
South Africa}
\end{center}
\vskip .5in
\begin{center}
{\bf ABSTRACT}
\end{center}

\begin{quotation}\noindent
\baselineskip 18pt
We construct a class of operators, given by Schur polynomials, in ABJM theory.  By computing two point functions at finite $N$
we confirm these are diagonal for this class of operators in the free field limit. We also calculate exact three and multi point
correlators in the zero coupling limit. Finally, we consider a particular nontrivial background produced by an operator with an $R$-charge
of $O(N^2)$. We show that the nonplanar corrections (which can no longer be neglected, even at large $N$) can be resummed to give
a $1/(N+M)$ expansion for correlators computed in this background.
\end{quotation}
\vskip .6in
\hspace{.3in}May 2011\\
\end{titlepage}
\vfill
\section{Introduction:}
\baselineskip 19pt
\noindent
Recently Aharony, Bergman, Jafferis and Maldacena (ABJM) proposed a gauge theory dual to M theory on $AdS_4\times S^7/Z_k$ with $N$ units of four
form flux \cite{Abjm}(see also \cite{Benna,Bhat,Bandres,Abj} for further results). This gauge theory is a three dimensional Chern-Simons-matter
theory with gauge group $U(N)\times U(N)$ or $SU(N) \times SU(N)$ with the Chern Simons levels $k$ and $-k$ associated with the each gauge groups
respectively. Further this theory has explicit ${\cal N}=6$ superconformal symmetry and the theory consists of four complex scalar fields
$A_i = (A_1, A_2) \;\;{\rm and}\;\; B_i^\dagger = (B_1^\dagger, B_2^\dagger)$. Fields $A$ and $B^\dagger$ both have conformal dimension $1/2$ and
carry $1/2$ unit of $R$-charge. These two pairs of fields transform in bifundamental representations of the gauge group. $A'$s transform in the
$(N, \bar N)$, while $B^{\dagger'}$s in the $( \bar N ,N)$. This theory enjoys a sensible large $N$ limit like ${\cal N}= 4$ SYM theory. One can
define a 't Hooft coupling constant by $\lambda=\frac{N}{k}$. Thus for $k >> N$, the theory is in the weak coupling limit. The theory goes
to the strong coupling regime for $k << N$. The corresponding gravitational dual is either M theory on $AdS_4 \times S^7/Z_k$ for
$k << N^{\frac{1}{5}}$ or type IIA theory on $AdS_4 \times CP^3$ for $N >> k >> N^{\frac{1}{5}}$.\\

\noindent
In the setting of ABJM theory, Berenstein and Transcanelli have initiated the study of
the M theory geometry using a field theory analysis \cite{Beren}(see also \cite{Sheikh,Beren1} for similar type of study). They considered the free field
regime of the ABJM theory on $R \times S^2$ and discussed half-BPS operators and their description in terms of droplets of free fermions and
Young tableaux. Their operators are labeled by Young tableaux with a maximum of $N$ rows. The number of boxes in the Young tableaux is equal to the
$R$-charge of the corresponding operator. These BPS operators are well described by single trace operators when one considers $R$-charges less than
$N$. They also found dual gravity descriptions of these operators in
terms of giant gravitons. These giants are given by either M2-brane growing in the $AdS_4$ or M5-brane wrapping a submanifold of $S^7$ depending
on the symmetric or antisymmetric representations of the operators for $k << N^{\frac{1}{5}}$. However for $N >> k >> N^{\frac{1}{5}}$, the M2-brane
is replaced by a D2-brane growing into $AdS_4$ and instead of M5-brane, D4-brane wrapping on some circle of $CP^3$. In the context of $AdS_5/CFT_4$ \cite{Malda,Gub,Wit} correspondence, the giant gravitons were first studied in \cite{Herzog} and  same are constructed in \cite{Nishioka} for ABJM theory. These giant gravitons has been further studied in \cite{Hamilton,Hamilton1,Murugan}.\\

\noindent
In ${\cal N}= 4$ SYM theory, we know that trace operators do not provide a useful basis for gauge invariant operators, when these operators have large
$R$-charge \cite{Bala}. It is natural to expect the same is true for the AMJM theory. Therefore in this paper, we first consider the trace operators
for large $R$-charge in ABJM theory and find they are not a useful set of operators to consider. We then propose a class of Schur polynomials for this
theory for large $R$-charge. Although they do not provide a complete basis for the gauge invariant operators, these polynomials diagonalize the two
point function in the free field limit. Our work thus provides a useful first step which might be extended to find a complete basis. The operators
we study are already very useful and we demonstrate how to construct the large $N$ expansion of this theory for trivial and nontrivial background.\\

\noindent
This paper is organized as follows: in section \ref{review} we review some basic facts of half-BPS operators of ${\cal N}=4$ SYM theory and group theory.
Then in section \ref{schur} we propose Schur polynomials for ABJM theory. In section \ref{two} and \ref{three} we find out the two, three and multi point
functions for the proposed Schur polynomials. We compare results of these correlators with the correlators of ${\cal N}=4$ SYM theory in section
\ref{summary}. In section \ref{amply}, we compute the amplitude of multi trace operators with and without nontrivial
back ground. Finally we conclude our results in section \ref{conc}.

\section{Review and Notation:}\label{review}
\noindent
In this section we recall some basic results concerned with half-BPS operators in ${\cal N} =4$ SYM gauge theory, group theory and notation which we
use in the rest of the paper. For details see these papers \cite{Guna,Andri,Ahar,Skiba,Andri1,Koch,Koch1,Koch2} and references their in. We mainly
follow the paper \cite{Sanjay,Corley} and use their notation.
\subsection{\it Half-BPS operators in ${\cal N} =4$ SYM:}
In AdS/CFT correspondence the operators of ${\cal N} =4$ SYM theory on $R \times S^3$ dual to states of the string theory on
curved $AdS_5 \times S^5$. Gauge theory operators are constructed from the combinations of three complex scalar fields and their conjugates. These
fields are complex linear combinations of six real scalar fields $\phi_i$ in the adjoint representation of the $U(N)$. We group these complex scalar
fields as $$ Z= \phi_1 + i\phi_2,\: Y = \phi_3 + i\phi_4 \quad{\rm{and}}\quad X= \phi_5 + i\phi_6.$$ Under the $U(N)$ gauge group these complex
scalars transform as $Z\rightarrow U^\dagger Z U$. The half-BPS operators are constructed from one type of complex scalar and the simplest
half-BPS operators are just trace operators of the form $\prod_{n_i}\big[\Tr(Z^l)\big]^{n_i}$. Here $l$ counts the $R$-charge of the operator.
In the standard AdS/CFT map, every trace operator correspond to a single particle in the dual AdS space, double trace operator with the two
particle state and so on. However, the orthogonality condition of states created by operators with different number of traces implies that
the description in terms of trace operator is valid only when the $R$-charge $l$ of the operator is less than $\sqrt N$ \cite{Bala}. Beyond this
limit, we need Schur polynomial, which furnish the correct orthogonality properties\cite{Sanjay}. The form of this Schur polynomial is
\begin{eqnarray}
\nonumber\chi_R(Z)&=& \frac{1}{n!}\sum_{\sigma \in S_n}\chi_R(\sigma)\Tr(\sigma Z)\nonumber
\label{schurn4}
\end{eqnarray}
where
\begin{equation}
\Tr(\sigma Z)\equiv \sum_{i_1,i_2\cdots i_n}Z_{i_{\sigma(1)}}^{i_1}Z_{i_{\sigma(2)}}^{i_2}\cdots Z_{i_{\sigma(n)}}^{i_n}.
\label{schurn41}
\end{equation}
$R$ is the representation of a specific Young diagram with $n$ boxes. This Young diagram labels both a representation of $U(N)$ and a representation of $S_n$. $\chi_R(\sigma)$ is the character or trace of the element $\sigma \in S_n$ in the representation $R.$\\
\noindent
When we say two operators are orthogonal, we mean their two point function vanishes. For half-BPS operators the exact two point function is obtained
in the free field limit. To compute the two point function we use the basic Wick contraction between two fields
$$\big\langle Z_{ij}(x)Z_{kl}^\dagger(y)\big\rangle = \big\langle Y_{ij}(x)Y_{kl}^\dagger(y)\big\rangle = \big\langle X_{ij}(x)X_{kl}^\dagger(y)\big\rangle= \frac{\delta_{il}\delta_{jk}}{(y - x)^2}.$$
The space time dependence of the correlators we consider is trivial. The nontrivial contribution to the correlators comes from the factor obtained by
performing the sum over $U(N)$ indices. We thus often suppress the space time dependence. However, we can easily bring back this dependence at any stage
of the calculation. The sum over $U(N)$ indices frequently takes the form
$$\sum_{i_1,i_2\cdots i_n}\delta_{i_{\sigma(1)}}^{i_1}\delta_{i_{\sigma(2)}}^{i_2}\cdots \delta_{i_{\sigma(n)}}^{i_n} = N^{C(\sigma)}$$
where each index $i_1,\cdots i_n$ takes an integer value from 1 to $N$, $\sigma$ is the permutation element and $ C(\sigma)$ is the number of cycles
in the permutation $\sigma$.\\
\noindent
For the simplicity of calculation, we introduce multi index notation $I(n)$ which is a shorthand for a set of indices, instead of writing out strings
of delta functions carrying $n$ different indices. In this multi index notation the sum can be reduces to
\begin{equation}
\sum_{i_1,i_2\cdots i_n}\delta_{i_{\sigma(1)}}^{i_1}\delta_{i_{\sigma(2)}}^{i_2}\cdots \delta_{i_{\sigma(n)}}^{i_n} = \sum_I \delta\Bigg(\stackrel{I(n)}{I\big(\sigma(n)\big)}\Bigg)=N^{C(\sigma)}.
\label{multiI}
\end{equation}
For later use we write down the trace part of the Schur polynomial of eqn.(\ref{schurn41}) in the language of multi index notation which is
\begin{eqnarray}
\nonumber\Tr(\sigma Z)&\equiv& \hspace{-.1in}\sum_{i_1,i_2\cdots i_n}Z_{i_{\sigma(1)}}^{i_1}Z_{i_{\sigma(2)}}^{i_2}\cdots Z_{i_{\sigma(n)}}^{i_n}\\
&=&\sum_I Z\Bigg(\stackrel{I(n)}{I\big(\sigma(n)\big)}\Bigg).
\label{schurn42}
\end{eqnarray}
Having this brief review on the ${\cal N} =4$ SYM theory, we will now review a few background facts from group theory.
\subsection{\it Groups:}
We start this subsection by  making the comment on Schur polynomials that these are the characters of the unitary group in their irreducible representations that means,
\begin{equation}
\chi_R(U)= \frac{1}{n!}\sum_{\sigma \in S_n}\chi_R(\sigma)\Tr(\sigma U).\nonumber
\end{equation}
By considering $U=1$, we derive the expression for the dimension of a representation of the unitary group as
\begin{equation}
Dim_N(R)= \frac{1}{n!}\sum_{\sigma \in S_n}\chi_R(\sigma)N^{C(\sigma)}.
\label{Dim}
\end{equation}
We can evaluate the value of this dimension of a representation from the Young diagram by using the formula
\begin{equation}
Dim_N(R)=\prod_{i,j}\frac{(N-i+j)}{h_{i,j}}.
\label{dim1}
\end{equation}
Here $i$ and $j$ label the rows and columns of the diagram respectably. $h_{i,j}$ is the hook number of each box of the diagram.\\
\noindent
We can also evaluate $d_R$ the dimension of a representation $R$ of the permutation group $S_n$ from corresponding Young diagram by using the formula
\begin{equation}
d_R=\frac{n!}{\prod_{i,j}h_{i,j}}.
\label{dim2}
\end{equation}
From eqns.(\ref{dim1}) and (\ref{dim2}) one can easily establish the relation of the product of the weights of the Young diagram with dimension of the
representation in the following way
\begin{equation}
f_R =\prod_{i,j}(N-i+j)= \frac{n! Dim_N(R)}{d_R}.
\label{weight}
\end{equation}

\noindent
Another useful fact in group theory is that if the character, in an irreducible representation of the symmetric group $R$, of a product of an element $C$ of the group algebra which commutes with everything with an arbitrary element $\sigma$. Then the character can be expanded in to a product of characters as follows;
\begin{equation}
\chi_R(C \sigma)=\frac{\chi_R(C)\chi_R(\sigma)}{d_R}.
\label{prodch}
\end{equation}
The element $C$ has to be either averages over the symmetric group of the form $\sum_{\alpha,\rho}f(\alpha\rho\alpha^{-1})\rho$ or $\sum_{\rho}g(\rho)\rho$ where $g(\rho)$ is a class function.\\

\noindent
With these we should mention two orthogonality relations of characters
\begin{equation}
\hspace{-.2in}\sum_{\sigma\in S_n} \chi_R(\sigma)\chi_S(\sigma^{-1})= n!\delta_{RS}
\label{ortho1}
\end{equation}
and
\begin{equation}
\sum_{\sigma\in S_n} \chi_R(\sigma\alpha)\chi_S(\sigma^{-1})= \frac{n!\delta_{RS}}{d_R}\chi_R(\alpha).
\label{ortho2}
\end{equation}
We close this section with a discussion of the product rule for Schur polynomials. Start with the three irreducible representation $R_1, R_2$ and $S$,
having $n_{R_1}, n_{R_2}$ and $n_S$ boxes in their respective Young diagrams, so that $n_S = n_{R_1} + n_{R_2}$. The product rule says the
product of Schur polynomials of irreducible representation $R_1$ and $R_2$ can be written as
\begin{equation}
\chi_{R_1}(Z)\chi_{R_2}(Z)=\sum_S g(R_1,R_2;S)\chi_{S}(Z).
\label{product1}
\end{equation}
The Littlewood-Richardson number $g(R_1,R_2;S)$ counts the number of times irreducible $U(N)$ representation $S$ appears in the direct product of
irreducible $U(N)$ representations $R_1$ and $R_2$. By repeated use of this product rule we can write the direct product of
$\chi_{R_1}(Z)\chi_{R_2}(Z)\cdots\chi_{R_l}(Z)$ as
\begin{eqnarray}
\nonumber\prod_{i=1}^l\chi_{R_i}(Z) &=& \hspace{-.2in}\sum_{S_1,S_2\cdots S_{l-2},S} \hspace{-.05in}
g(R_1,R_2;S_1)g(S_1,R_3;S_2)\cdots g(S_{l-2},R_l;S)\chi_{S}(Z)\\
&=&\hspace{.05in}\sum_S g(R_1,R_2\cdots R_l;S)\chi_{S}(Z).
\label{product2}
\end{eqnarray}

\section{Schur Polynomials for ABJM:}\label{schur}
\noindent
The main goal of this section is to construct a class of gauge invariant operators for ABJM gauge theory. In particular we are interested
in constructing half-BPS operators for large $R$-charge that is, operators with an R-charge which depends on $N$. In the ABJM theory, the
gauge group is $U(N)\times U(N)$ and it consists of four complex scalar fields $A_i=(A_1, A_2)$ and $B_i^\dagger = (B_1^\dagger, B_2^\dagger)$.
We distinguish  the first and second gauge group by using the notation $U_1(N)\times U_2(N)$. Under this group $A$ and $B^\dagger$ transforms
as $$A\longrightarrow U_1^\dagger A U_2 \;\; {\rm and}\;\; B^\dagger\longrightarrow U_2^\dagger B^\dagger U_1.$$ Therefore in the matrix
notation we can write $A$ and $B^\dagger$ in the following way $$A_j^i \;\; {\rm and} \;\; (B^\dagger)_i^j$$ where $i$ and $j$ are gauge
indices of $U_1$ and $U_2$. In order to get a gauge invariant operator, we need to contract $U_1$ and $U_2$ indices. We can thus have the
combinations of either $(AA^\dagger)$ or $(BB^\dagger)$ or $(AB^\dagger)$ or $(A^\dagger B)$. For the first two combinations $R$-charge is zero but
its $1$ for the last two options with conformal dimension $1$ for all combinations. However the BPS inequality demands that $R$-charge should
be less than or equal to the conformal dimension. Therefore we can only construct the half-BPS operators from the combination of either
$(AB^\dagger)$ or $(A^\dagger B)$ which saturate the BPS bound where conformal dimension is equal to $R$-charge. If we see this object as an $N \times N$ matrix, the indices are contracted by taking traces.  Therefore,
by using our experience of ${\cal N} =4$ SYM theory, we can write down the simplest gauge invariant half-BPS operators for ABJM theory of
the form $$\prod_{n_i}[\Tr((AB^\dagger)^l)]^{n_i}.$$ In these half-BPS operators the complex scalar $A$ has to be either $A_1$ or $A_2$ and
similarly for complex scalar $B^\dagger$. According to \cite{Beren,Sheikh} these operators are represented by Young tableaux of boxes equal to
number of $(AB^\dagger)$ fields and at most $N$ rows. If these operators are the correct gauge invariant operators to study the theory, then
they should satisfy the orthogonality condition. To check the orthogonality, following the paper \cite{Bala}, we consider these two operators
\begin{eqnarray}
\nonumber {\cal{O}}_1 &=& \Tr\big((AB^\dagger)^l\big)\\\nonumber
{\rm and}\quad {\cal{O}}_2 &=& \Tr\big((AB^\dagger)^{l_1}\big)\Tr\big((AB^\dagger)^{l_2}\big) \hspace{.4in}{\rm with}\hspace{.4in}l_1+l_2=l.
\end{eqnarray}
 We then compute following two point functions in the free field limit using free field Wick contractions. In the leading order these are
\begin{eqnarray}
\nonumber\Big\langle{\cal{O}}_1{\cal{O}}_1^\dagger\Big\rangle &\sim& l N^{2l}
\end{eqnarray}
and
\begin{eqnarray}
\nonumber\Big\langle{\cal{O}}_2{\cal{O}}_2^\dagger\Big\rangle &\sim& l_1 l_2 N^{2l}.
\end{eqnarray}
In this computation, we drop the space time dependence. With these results, we compute the normalized 2-point function of two different operators
again in the leading order. The result is
\begin{equation}
\frac{\Big\langle{\cal{O}}_1{\cal{O}}_2^\dagger\Big\rangle}{\sqrt{\Big\langle{\cal{O}}_1
{\cal{O}}_1^\dagger\Big\rangle}\sqrt{\Big\langle{\cal{O}}_2
{\cal{O}}_2^\dagger\Big\rangle}} \sim \frac{\sqrt{l l_1 l_2}}{N}.
\end{equation}
This implies that at the large $N$ the operators ${\cal{O}}_1$ and ${\cal{O}}_2$ are orthogonal provided the factor $\sqrt{l l_1 l_2}$ is much less than
$N$. Therefore if $R$-charge of the operators $l$ is less than $N^{2/3}$, trace operators can be used to study the theory. However, for large $R$-charge
we need a different type of operator to replace the trace operators. Experience gained from ${\cal N} =4$ SYM theory, suggests that a suitable set of
orthogonal operators are provided by the Schur polynomials. For a gauge group $U(N)\times U(N)$ we seek to generalize this result. In particular we
construct polynomials of trace operators of which generalize the Schur polynomials of ${\cal N} =4$ SYM theory. Therefore we can guess the simplest
operator should look like
\begin{equation}
\chi_R(AB^\dagger)=\frac{1}{n!}\sum_{\sigma\in S_n}\chi_R(\sigma)\Tr(\sigma(AB^\dagger)).
\end{equation}
However this guess will be correct if this Schur polynomial satisfy the orthogonality condition. To check that we compute the two point function in the free field limit in the next section. \\

\noindent
Before going to calculate the two point function, for later use we would like to do this little exercise.
\begin{eqnarray}
\nonumber && \sum_{\sigma,\rho \in S_n}\chi_R(\sigma)\chi_S(\rho)\Tr((\sigma A)(\rho B^{\dagger}))\\\nonumber &=& \sum_{\sigma,\rho \in S_n}\chi_R(\sigma)\chi_S(\rho)A_{j_{\sigma(1)}}^{i_1}\cdots A_{j_{\sigma(n)}}^{i_n}B_{i_{\rho(1)}}^{\dagger j_1}\cdots B_{i_{\rho(n)}}^{\dagger j_n}\\\nonumber
&=& \sum_{\sigma,\rho \in S_n}\chi_R(\sigma)\chi_S(\rho)A_{j_{\sigma(1)}}^{i_{\rho^{-1}(1)}}\cdots A_{j_{\sigma(n)}}^{i_{\rho^{-1}(n)}}B_{i_{1}}^{\dagger j_1}\cdots B_{i_{n}}^{\dagger j_n}\\\nonumber
&=& \sum_{\sigma,\rho \in S_n}\chi_R(\sigma)\chi_S(\rho)A_{j_{\sigma(\psi(1))}}^{i_{\rho^{-1}(\psi(1))}}\cdots A_{j_{\sigma(\psi(n))}}^{i_{\rho^{-1}(\psi(n))}}B_{i_{1}}^{\dagger j_1}\cdots B_{i_{n}}^{\dagger j_n}.
\end{eqnarray}
By considering $\psi = \rho$, the form reduces to
\begin{eqnarray}
\nonumber&&\sum_{\sigma,\rho \in S_n}\chi_R(\sigma)\chi_S(\rho)A_{j_{\sigma(\rho(1))}}^{i_{1}}\cdots A_{j_{\sigma(\rho(n))}}^{i_{n}}B_{i_{1}}^{\dagger j_1}\cdots B_{i_{n}}^{\dagger j_n}.
\end{eqnarray}
Again consider $\tau = \sigma\rho \Rightarrow \sigma = \tau\rho^{-1}$ and the above form recast as
\begin{eqnarray}
\nonumber&& \sum_{\tau,\rho \in S_n}\chi_R(\tau\rho^{-1})\chi_S(\rho)A_{j_{\tau(1)}}^{i_{1}}\cdots A_{j_{\tau(n)}}^{i_{n}}B_{i_{1}}^{\dagger j_1}\cdots B_{i_{n}}^{\dagger j_n}.
\end{eqnarray}
Using eqn.(\ref{ortho2}) we can reformulate the above expression like
\begin{eqnarray}
\nonumber&& \delta_{RS}\frac{n!}{d_R}\sum_{\tau\in S_n}\chi_R(\tau)(AB^{\dagger})_{j_{\tau(1)}}^{j_{1}}\cdots (AB^{\dagger})_{j_{\tau(n)}}^{j_{n}}\\\nonumber
&=& \delta_{RS}\frac{(n!)^2}{d_R}\frac{1}{n!}\sum_{\tau\in S_n}\chi_R(\tau)\Tr(\tau (AB^{\dagger}))\\\nonumber
&=& \delta_{RS}\frac{(n!)^2}{d_R}\chi_R(AB^{\dagger}).
\end{eqnarray}
Thus our Schur polynomial can be rewritten as follows
\begin{equation}
\chi_R(AB^{\dagger})=\delta_{RS}\frac{d_R}{(n!)^2}\sum_{\sigma,\rho \in S_n}\chi_R(\sigma)\chi_S(\rho)\Tr((\sigma A)(\rho B^{\dagger}))
\label{schurabjm1}
\end{equation}
and
\begin{equation}
\chi_R(A^{\dagger}B)= \delta_{RS}\frac{d_R}{(n!)^2}\sum_{\sigma,\rho \in S_n}\chi_R(\sigma)\chi_S(\rho)\Tr((\sigma A^{\dagger})(\rho B)).
\label{schurabjm2}
\end{equation}

\section{Two point function:}\label{two}
In this section we compute the two point function of our proposed Schur polynomials. The two point function of our interest is
\begin{equation}
\bigg\langle\chi_R(AB^{\dagger})\chi_S(A^{\dagger}B)\bigg\rangle.
\end{equation}
The representation $R$ of symmetric group $S_{n_R}$ has $n_R$ boxes in the Young diagram and the representation $S$ of symmetric group $S_{n_S}$ has
$n_S$ boxes. In the calculation we suppress the space time dependence but when we wish to bring back the space time dependence, these two representation
will be in at different points in the space time. For computational simplicity, it is convenient to use the above two relations of eqn.(\ref{schurabjm1})
and (\ref{schurabjm2}) for these two Schur polynomials and get
\begin{equation}
\frac{d_R d_S}{(n_R!n_S!)^2}\sum_{\sigma,\rho, \tau, \gamma}\chi_R(\sigma)\chi_R(\rho)\chi_S(\tau)\chi_S(\gamma)\bigg\langle\Tr((\sigma A)(\rho B^{\dagger}))\Tr((\tau A^{\dagger})(\gamma B))\bigg\rangle.
\end{equation}
Here $\sigma, \rho$ are the elements of $S_{n_R}$ and $\tau, \gamma$ are the elements of $S_{n_S}$. We now use the multi index form of trace part which
is introduced in eqn.(\ref{schurn42}) and we have
\begin{eqnarray}
&& \frac{d_R d_S}{(n_R!n_S!)^2}\sum_{\stackrel{\sigma,\rho, \tau, \gamma}{I,J,M,N}}\chi_R(\sigma)\chi_R(\rho)\chi_S(\tau)\chi_S(\gamma)\times\\\nonumber
&&\hspace{1in}\Bigg\langle A\Bigg(\stackrel{I(n_R)}{J(\sigma(n_R))}\Bigg) B^{\dagger}\Bigg(\stackrel{J(n_R)}{I(\rho(n_R))}\Bigg)\times\\\nonumber
&&\hspace{1in}A^{\dagger}\Bigg(\stackrel{M(n_S)}{N(\tau(n_S))}\Bigg)B\Bigg(\stackrel{N(n_S)}{M(\gamma(n_S))}\Bigg)
\Bigg\rangle.
\end{eqnarray}
By performing the Wick contractions between fields $A, A^\dagger$ and $B^\dagger, B$ we gain two extra sum over permutations $\alpha$ and $\beta$. Both
of them belong to the symmetric group $S_{n_S}$. As a result the two point function appears in the following way
\begin{eqnarray}
&& \frac{d_R d_S}{(n_R!n_S!)^2}\sum_{\stackrel{\sigma,\rho, \tau, \gamma ,\alpha,\beta}{I,J,M,N}}\chi_R(\sigma)\chi_R(\rho)\chi_S(\tau)\chi_S(\gamma)\times\\\nonumber
&&\hspace{1in} \delta\Bigg(\stackrel{I(n_R)}{N(\alpha\tau(n_S))}\Bigg)
\delta\Bigg(\stackrel{M(\alpha(n_S))}{J(\sigma(n_R))}\Bigg)\times\\\nonumber
&&\hspace{1in}\delta\Bigg(\stackrel{J(n_R)}{M(\beta\gamma(n_S))}\Bigg)
\delta\Bigg(\stackrel{N(\beta(n_S))}{I(\rho(n_R))}\Bigg)\\\nonumber
&=& \frac{d_R d_S}{(n_R!n_S!)^2}\sum_{\stackrel{\sigma,\rho, \tau, \gamma ,\alpha,\beta}{I,J,M,N}}\chi_R(\sigma)\chi_R(\rho)\chi_S(\tau)\chi_S(\gamma)\times\\\nonumber
&&\hspace{.9in} \delta\Bigg(\stackrel{I(\tau^{-1}\alpha^{-1}(n_R))}{N(n_S)}\Bigg)
\delta\Bigg(\stackrel{M(n_S)}{J(\alpha^{-1}\sigma(n_R))}\Bigg)\times\\\nonumber
&&\hspace{.9in} \delta\Bigg(\stackrel{J(\gamma^{-1}\beta^{-1}(n_R))}{M(n_S)}\Bigg)
\delta\Bigg(\stackrel{N(n_S)}{I(\beta^{-1}\rho(n_R))}\Bigg).
\end{eqnarray}
After doing the sum over $M$ and $N$ single multi index we left with
\begin{eqnarray}
&& \frac{d_R d_S}{(n_R!n_S!)^2}\sum_{\stackrel{\sigma,\rho, \tau, \gamma, \alpha,\beta}{I,J}}\chi_R(\sigma)\chi_R(\rho)\chi_S(\tau)\chi_S(\gamma)\times\\\nonumber
&&\hspace{1in} \delta\Bigg(\stackrel{I(\tau^{-1}\alpha^{-1}(n_R))}{I(\beta^{-1}\rho(n_R))}\Bigg)
\delta\Bigg(\stackrel{J(\gamma^{-1}\beta^{-1}(n_R))}{J(\alpha^{-1}\sigma(n_R))}\Bigg)\\\nonumber
&=& \frac{d_R d_S}{(n_R!n_S!)^2}\sum_{\stackrel{\sigma,\rho, \tau, \gamma, \alpha,\beta}{I,J}}\chi_R(\sigma)\chi_R(\rho)\chi_S(\tau)\chi_S(\gamma)\times\\\nonumber
&&\hspace{.8in} \delta\Bigg(\stackrel{I(\rho^{-1}\beta\tau^{-1}\alpha^{-1}(n_R))}{I(n_R)}\Bigg)
\delta\Bigg(\stackrel{J(\sigma^{-1}\alpha \gamma^{-1}\beta^{-1}(n_R))}{J(n_R)}\Bigg).
\end{eqnarray}
We replace the sum of free field contractions by using the eqn.(\ref{multiI}) to obtain the  form
\begin{equation}
\frac{d_R d_S}{(n_R!n_S!)^2}\sum_{\sigma,\rho, \tau, \gamma, \alpha,\beta}\chi_R(\sigma)\chi_R(\rho)\chi_S(\tau)\chi_S(\gamma) N^{C_1(\rho^{-1}\beta\tau^{-1}\alpha^{-1})} N^{C_2(\sigma^{-1}\alpha \gamma^{-1}\beta^{-1})}.
\end{equation}
Purely for computation purpose, we introduce two extra summed permutations $p$ and $q$ those are constrained by two delta functions to simplify the
exponents of $N$ and get the first line. These delta functions are $1$ when
arguments are identity and $0$ otherwise. We then do the sum over $\tau$ and $\gamma$ to obtain the second line
\begin{eqnarray}
\nonumber&& \frac{d_R d_S}{(n_R!n_S!)^2}\sum_{\sigma,\rho, \tau, \gamma, \alpha,\beta, p,q}\chi_R(\sigma)\chi_R(\rho)\chi_S(\tau)\chi_S(\gamma)N^{C_1(p)}N^{C_2(q)}
\delta(p^{-1}\rho^{-1}\beta\tau^{-1}\alpha^{-1})\times\\
&&\hspace{1.5in}\delta(q^{-1}\sigma^{-1}\alpha \gamma^{-1}\beta^{-1})\\\nonumber
&=& \frac{d_R d_S}{(n_R!n_S!)^2}\sum_{\sigma,\rho, p, q, \alpha,\beta }\chi_R(\sigma)\chi_R(\rho)\chi_S(\beta\alpha^{-1}p^{-1}\rho^{-1})\chi_S(\alpha\beta^{-1}q^{-1}
\sigma^{-1})N^{C_1(p)}N^{C_2(q)}.
\end{eqnarray}
Since $\sum_p N^{C_1(p)}p$ and $\sum_q N^{C_2(q)}q$ commute with the any element of $S_{n_S}$, following eqn.(\ref{prodch}), we can expand the last two
characters into  product of characters. Then use the eqn.(\ref{ortho2}) to recast the above form as
\begin{eqnarray}
&& \frac{d_R d_S}{(n_R!n_S!)^2}\sum_{\sigma,\rho, p, q, \alpha,\beta}\chi_R(\sigma)\chi_R(\rho)
\chi_S(p^{-1})\frac{1}{d_S}\chi_S(\beta\alpha^{-1})
\frac{1}{d_S}\chi_S(\rho^{-1})\times\\\nonumber
&&\hspace{1.2in}\chi_S(q^{-1})\frac{1}{d_S}\chi_S(\alpha\beta^{-1})
\frac{1}{d_S}\chi_S(\sigma^{-1})N^{C_1(p)}N^{C_2(q)}.
\end{eqnarray}
Now by doing the sum over $\alpha$ and $\beta$ we gain the extra factor $n_S!^2$. Performing the sum over $\sigma$ and $\rho$ we also get a delta
function $\delta_{RS}$ and a factor $n_R!^2$. Therefore, by considering all the results together we can write the above expression like
\begin{eqnarray}
\nonumber&&\sum_{ p, q }\delta_{RS}\frac{\chi_S(p^{-1})\chi_S(q^{-1})}{d_S^2}N^{C_1(p)}N^{C_2(q)}\\
&=&\Big( n!\frac{Dim_N(S)}{d_S}\Big)^2\delta_{RS} = f_S^2\;\delta_{RS}.
\label{twpr}
\end{eqnarray}
To get the last line we have used the eqn.(\ref{Dim}) and (\ref{weight}). Due the presence of delta function its clear that the two point function will
only be non zero when both the representation $R$ and $S$ are exactly same. Therefore our proposed Schur polynomial satisfy the orthogonality condition
and we can demand these Schur polynomials are the correct gauge invariant operators to study the ABJM theory for large $R$-charge.\\

\noindent
With this discussion on two point function we move to compute the three and multi point functions in the next section.

\section{Three and Multi point function:}\label{three}
\subsection{\it Three point function:}
Here we consider the following three point function of Schur polynomials.
\begin{equation}
\Bigg\langle\chi_{R_1}(AB^{\dagger})\chi_{R_2}(AB^{\dagger})\chi_S(A^{\dagger}B)\Bigg\rangle
\end{equation}
When we recall the space time coordinate dependence, all Schur polynomials will be in at three different points. Representation
$R_1$ has $n_{R_1}$ boxes, $R_2$ has $n_{R_2}$ boxes, $S$ has $n_{S}$ boxes and $n_{S}= n_{R_1}+n_{R_2}$. Two compute this three point
function we use the product rule of eqn.(\ref{product1}) following the logic of \cite{Koch3}. Once we use the product rule on $\chi_{R_1}(AB^{\dagger})$ and $\chi_{R_2}(AB^{\dagger})$
the three point function reduces to
\begin{eqnarray}
\nonumber&&\sum_{S^{\prime}}g(R_1,R_2;S^{\prime})\Bigg\langle\chi_{S^{\prime}}(AB^{\dagger})
\chi_S(A^{\dagger}B)\Bigg\rangle.
\end{eqnarray}
Now we can easily use the result of two point function which we have computed in the previous section and the result of the three point function becomes
\begin{eqnarray}
&&\sum_{S^{\prime}}g(R_1,R_2;S^{\prime})f_S^2\;\delta_{S^\prime S}=g(R_1,R_2;S)f_S^2
\end{eqnarray}

\subsection{\it Multi point function:}
\noindent
The correlation function of our interest is:
\begin{equation}
\Bigg\langle\chi_{R_1}(AB^{\dagger})\chi_{R_2}(AB^{\dagger})\cdot\cdot\cdot \chi_{R_l}(AB^{\dagger}) \chi_{S_1}(A^{\dagger}B)\chi_{S_2}(A^{\dagger}B)\cdots\chi_{S_k}(A^{\dagger}B)  \Bigg\rangle.
\end{equation}
Representation
$R_i$ has $n_{R_i}$, $S_j$ has $n_{S_j}$ boxes and  $\sum_{i=1}^l n_{R_i}=\sum_{j=1}^k n_{S_j}$. As earlier if we want to introduce the space time coordinate dependence, all Schur polynomials will be in at different points. To calculate this multi
point function we again use the product rule of eqn.(\ref{product2}) and then in the second line we use the result of two point function of
eqn.(\ref{twpr}) and obtain the form as
 \begin{eqnarray}
\nonumber&&\sum_{S_1^\prime,S_2^\prime\cdots S_{l-2}^\prime,S^\prime} g(R_1,R_2;S_1^\prime)g(S_1^\prime,R_3;S_2^\prime)\cdots g(S_{l-2}^\prime,R_l;S^\prime)\Bigg\langle\chi_{S^\prime}(AB^{\dagger})
\chi_{S^{\prime\prime}}(A^{\dagger}B)\Bigg\rangle \times\\\nonumber
&&\sum_{S_1^{\prime\prime},S_2^{\prime\prime}\cdots S_{k-2}^{\prime\prime},S^{\prime\prime}} g(S_1,S_2;S_1^{\prime\prime})g(S_1^{\prime\prime},S_3;S_2^{\prime\prime})\cdots g(S_{k-2}^{\prime\prime},S_k;S^{\prime\prime})\\\nonumber
&=&\sum_{S_1^\prime,S_2^\prime\cdots S_{l-2}^\prime,S^\prime} g(R_1,R_2;S_1^\prime)g(S_1^\prime,R_3;S_2^\prime)\cdots g(S_{l-2}^\prime,R_l;S^\prime)f_{S^\prime}^2\;\delta_{S^\prime S^{\prime\prime}} \times\\\nonumber
&&\sum_{S_1^{\prime\prime},S_2^{\prime\prime}\cdots S_{k-2}^{\prime\prime},S^{\prime\prime}} g(S_1,S_2;S_1^{\prime\prime})g(S_1^{\prime\prime},S_3;S_2^{\prime\prime})\cdots g(S_{k-2}^{\prime\prime},S_k;S^{\prime\prime})\\\nonumber
&=&\sum_{S_1^\prime,S_2^\prime\cdots S_{l-2}^\prime,S^\prime} g(R_1,R_2;S_1^\prime)g(S_1^\prime,R_3;S_2^\prime)\cdots g(S_{l-2}^\prime,R_l;S^\prime)f_{S^\prime}^2\; \times\\\nonumber
&& \sum_{S_1^\prime,S_2^\prime\cdots S_{k-2}^\prime,S^\prime}g(S_1,S_2;S_1^{\prime})g(S_1^{\prime},S_3;S_2^{\prime})\cdots g(S_{k-2}^{\prime},S_k;S^{\prime})\\\nonumber
&&\\
&=&\sum_{S}g(R_1,R_2\cdots R_l;S)\,f_S^2\; g(S_1,S_2\cdots S_k;S).
\end{eqnarray}
In the last line we replace $S^\prime$ as $S$.
\section{Summary of results of ABJM correlators and comparison with the correlators of ${\cal N}=4 $ SYM:}\label{summary}
In this section we summaries the results of two, three and multi point functions for ABJM theory. In the summary of the results we bring back the space time dependence. In what follows, we first write down the results of ABJM theory and then the results of ${\cal N}=4 $ SYM theory. Finally compare the results of these two theories.\\
{\it \underline{ABJM}:}
For small $R$-charge we can compute the correlators without using Schur polynomials and the results are
\begin{eqnarray}
\nonumber\Big\langle{\cal{O}}_1(x){\cal{O}}_1^\dagger(y)\Big\rangle &\sim& \frac{l N^{2l}}{(y-x)^{4l}}
\end{eqnarray}
and
\begin{eqnarray}
\nonumber\Big\langle{\cal{O}}_2(x){\cal{O}}_2^\dagger(y)\Big\rangle &\sim& \frac{l_1 l_2 N^{2l}}{(y-x)^{4l}}.
\end{eqnarray}
However for the large $R$-charge we have to consider the Schur polynomials and the result of two point function is
\begin{eqnarray}
\nonumber\bigg\langle\chi_R(AB^{\dagger})(x)\chi_S(A^{\dagger}B)(y)\bigg\rangle&=&\Big( n!\frac{Dim_N(S)}{d_S}\Big)^2\frac{\delta_{RS}}{(y-x)^{4n_R}} = \frac{\delta_{RS}f_S^2}{(y-x)^{4n_R}}.
\end{eqnarray}
The three point function with large $R$-charge we find
\begin{eqnarray}
\nonumber&&\hspace{-.4in}\Bigg\langle\chi_{R_1}(AB^{\dagger})(x_1)\chi_{R_2}(AB^{\dagger})(x_2)\chi_S(A^{\dagger}B)(y)\Bigg\rangle
=\frac{g(R_1,R_2;S)\;f_S^2}{(y-x_1)^{4n_{R_1}}(y-x_2)^{4n_{R_2}}}.
\end{eqnarray}
To present the multi point function of the operators with large $R$-charge we consider only one spacial case where all operators represented by $S_j$ are located at the same space time point and the result we have
\begin{eqnarray}
\nonumber&&\hspace{-1.4in}\Bigg\langle\chi_{R_1}(AB^{\dagger})(x_1)\chi_{R_2}(AB^{\dagger})(x_2)\cdot\cdot\cdot \chi_{R_l}(AB^{\dagger})(x_l) \chi_{S_1}(A^{\dagger}B)(y)\chi_{S_2}(A^{\dagger}B)(y)\cdots\chi_{S_k}(A^{\dagger}B)(y)  \Bigg\rangle\\\nonumber
\hspace{.9in}&=&\sum_S \frac{g(R_1,R_2\cdots R_l;S)\,f_S^2\;  g(S_1,S_2\cdots S_k;S)}{(y-x_1)^{4n_{R_1}}\cdots(y-x_l)^{4n_{R_l}}}.
\end{eqnarray}
{\it \underline{${\cal N}=4 $ SYM}:}
The above all results  with the same condition are as follows.
The two point functions  with small $R$-charge are
\begin{eqnarray}
\nonumber\Big\langle{\cal{O}}_1(x){\cal{O}}_1^\dagger(y)\Big\rangle &\sim& \frac{l N^{l}}{(y-x)^{2l}}
\end{eqnarray}
and
\begin{eqnarray}
\nonumber\Big\langle{\cal{O}}_2(x){\cal{O}}_2^\dagger(y)\Big\rangle &\sim& \frac{l_1 l_2 N^{l}}{(y-x)^{2l}}.
\end{eqnarray}
The two point function with large $R$-charge
\begin{eqnarray}
\nonumber\bigg\langle\chi_R(Z)(x)\chi_S(Z^\dagger)(y)\bigg\rangle&=&\Big( n!\frac{Dim_N(S)}{d_S}\Big)\frac{\delta_{RS}}{(y-x)^{2n_R}} = \frac{\delta_{RS}f_S}{(y-x)^{2n_R}}.
\end{eqnarray}
The three point function of the operators with large $R$-charge
\begin{eqnarray}
\nonumber&&\hspace{-.4in}\Bigg\langle\chi_{R_1}(Z)(x_1)\chi_{R_2}(Z)(x_2)\chi_S(Z^\dagger)(y)\Bigg\rangle
=\frac{g(R_1,R_2;S)\;f_S}{(y-x_1)^{2n_{R_1}}(y-x_2)^{2n_{R_2}}}.
\end{eqnarray}
The multi point function with large $R$-charge
\begin{eqnarray}
\nonumber&&\hspace{-1.4in}\Bigg\langle\chi_{R_1}(Z)(x_1)\chi_{R_2}(Z)(x_2)\cdot\cdot\cdot \chi_{R_l}(Z)(x_l) \chi_{S_1}(Z^\dagger)(y)\chi_{S_2}(Z^\dagger)(y)\cdots\chi_{S_k}(Z^\dagger)(y)  \Bigg\rangle\\\nonumber
\hspace{.9in}&=&\sum_S \frac{g(R_1,R_2\cdots R_l;S)\,f_S\;  g(S_1,S_2\cdots S_k;S)}{(y-x_1)^{2n_{R_1}}\cdots(y-x_l)^{2n_{R_l}}}.
\end{eqnarray}
For small $R$-charge the $N$ dependence of the correlators comes directly, while for large $R$-charge the $N$ dependence appears from $f_S$ the product of weights of the Young diagram. The main difference of the results
between these two theories are for small $R$-charge, in the leading order the power of $N$ of ABJM is double than the power of $N$ of ${\cal N}=4 $ SYM theory and for large $R$-charge $f_S$ comes with single power in ${\cal N}=4 $ SYM while, it is squared in the ABJM theory. These differences are expected
because ABJM gauge theory is $U(N)\times U(N)$ whether ${\cal N}=4 $ SYM theory is only $U(N)$.\\

\noindent
From our experience of ${\cal N}=4 $ SYM theory, we know that large $N$ expansion $1/N$ is replaced by $1/(N+M)$ if the $R$-charge of the operator is order of $N^2$. Where $M$ is the number of columns in the representing Young diagram and its order of $N$. Thus one can immediately ask the question that is there any replacement of $1/N$ expansion for operator with $R$-charge of order $N^2$ in ABJM? To check that in the next section we compute the amplitudes of ABJM theory with and without presence of non trivial background.
\section{ABJM Amplitudes}\label{amply}
\noindent
Keep in mind the goal of our last paragraph, we would like to calculate the correlators of multi trace operators at zero coupling. This can be
computed easily by expressing the multi trace operators of interest in terms of Schur polynomials
\begin{equation}
\prod_i \Tr\Big((AB^{\dagger})^{n_i}\Big)= \sum_{R}\alpha_R \chi_R(AB^{\dagger}),\quad \quad\prod_j \Tr\Big((A^{\dagger}B)^{m_j}\Big)= \sum_{R}\beta_R \chi_R(A^{\dagger}B).
\end{equation}
Here the coefficients $\alpha_R$ and $\beta_R$ are independent of $N$. First we compute the correlator with a trivial background. Following \cite{Koch4,dey} we can easily write down the correlator as
\begin{eqnarray}
\nonumber{\cal{A}}\Big(\{n_i;m_j\},N\Big)\equiv \Bigg\langle\prod_{ij} \Tr\Big((AB^{\dagger})^{n_i}\Big)\Tr\Big((A^{\dagger}B)^{m_j}\Big)\Bigg\rangle &=& \sum_{R,S}\alpha_R\beta_S\bigg\langle \chi_R(AB^{\dagger})\chi_S(A^{\dagger}B)\bigg\rangle \\ &=&\sum_{R}\alpha_R\beta_R f_R^2.
\label{corrwob}
\end{eqnarray}

\begin{figure}
\begin{center}
\begin{psfrags}
\psfrag{A}[][]{\hspace{.3cm}N}
\psfrag{B}[][]{\hspace{.1cm}B}
\psfrag{C}[][]{\hspace{-.2cm}M}
\psfrag{D}[][]{$R$}
\psfrag{T}[][]{\hspace{-.5cm}=}
\psfrag{E}[][]{$+R$}
\psfrag{x}[][]{$\times$}
\includegraphics[width=13cm]{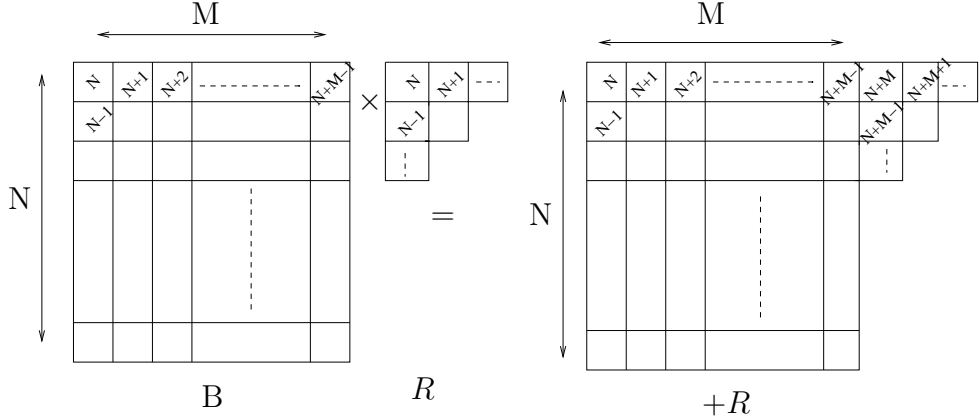}
\caption{Middle Young diagram represent operator $R$ which is multiplied by background diagram $B$ represented by left Young diagram and the right Young diagram $+R$ represent the product of these two operators.}
\label{backcom}
\end{psfrags}
\end{center}
\end{figure}

\noindent
Now we perform the calculation in presence of a non-trivial background operator denoted by $B$ depicted in fig.(\ref{backcom}). The background operator is represented by a rectangular Young tableau with $N$ rows and $M$ columns of order $N$. Every box of the tableau contains either $(AB^{\dagger})$ or $(A^{\dagger}B)$ field. We know the expectation value of an operator $\cal O$ in background $B$ is given by
\begin{equation}
\langle {\cal O} \rangle_B \equiv \frac{\Big\langle\chi_B(AB^{\dagger})\chi_B(A^{\dagger}B){\cal O}\Big\rangle}
{\Big\langle\chi_B(AB^{\dagger})\chi_B(A^{\dagger}B)\Big\rangle}.
\end{equation}

\noindent
Here the denominator plays the role of normalization factor. Therefore, the correlator in the background $B$ can be written as
\begin{eqnarray}
\nonumber{\cal{A}}_B\Big(\{n_i;m_j\},N\Big)&\equiv& \Bigg\langle\prod_{ij} \Tr\Big((AB^{\dagger})^{n_i}\Big)\Tr\Big((A^{\dagger}B)^{m_j}\Big)\Bigg\rangle_B \\ \nonumber
&=& \sum_{R,S}\alpha_R\beta_S\frac{\bigg\langle \chi_B(AB^{\dagger})\chi_R(AB^{\dagger})\chi_B(A^{\dagger}B)\chi_S(A^{\dagger}B)\bigg\rangle}{f_B^2} \\ \nonumber
&=&\sum_{R,S}\alpha_R\beta_S \frac{\bigg\langle\chi_{+R}(AB^{\dagger})\chi_{+S}(A^{\dagger}B)\bigg\rangle}{f_B^2}\\
&=&\sum_{R}\alpha_R\beta_R \Bigg(\frac{f_{+R}}{f_B}\Bigg)^2.
\label{corrwb}
\end{eqnarray}
\noindent
Here $f_{+R}$ is the product of the weights of the each box of Young tableau of $+R$ and recall that $f_B$ is the product of the weights of each box
of Young tableau $B$. All the weights of the box of $B$ is repeated in the $+R$. To get a clear idea see fig.(\ref{backcom}). Therefore, these weights
are canceled by the common weights of the $+R$ and product of the remaining weights of the $+R$ determine the ratio $\frac{f_{+R}}{f_B}$. Thus
$\frac{f_{+R}}{f_B}$ can be computed from  $f_R$ by just replacing $N\rightarrow N+M$. Comparing eqn.(\ref{corrwob}) and eqn.(\ref{corrwb}), we find
\begin{equation}
{\cal{A}}_B\Big(\{n_i;m_j\},N\Big)={\cal{A}}\Big(\{n_i;m_j\},N+M\Big).
\label{compare}
\end{equation}
\noindent
We know that the correlator $A({n_i;m_j},N)$ admits an expansion in $\frac{1}{N}$. Therefore from eqn.(\ref{compare}) we can demand that
$A_B({n_i;m_j},N)$ should admit an expansion in $\frac{1}{(N+M)}$ .

\section{Conclusion}\label{conc}
In this paper we first showed that trace operators do not provide a set of useful gauge invariant operators to study the ABJM gauge theory if
the $R$-charge of the operators is greater than $N^{2/3}$ as for ${\cal N}=4$ SYM theory. We thus need a different set of gauge invariant operators
to study this theory. Relying on experience gained from ${\cal N}=4$ SYM theory we propose a class of Schur polynomials which form a useful
set of operators in the large $R$-charge sector of ABJM theory. We have checked the orthogonality of our proposed polynomials by analysing two
point function. We also computed the exact three and multi point functions for these Schur polynomials.  Further, we calculated the amplitude of multi trace operators with and with out the presence of background operator of $R$-charge
of order $N^2$. These operators are dual to heavy objects in the string theory side and we can not neglect the back reaction on the dual geometry.
Due to this back reaction dual geometry will be modified and we have new geometry. Though this geometry is not explored in this paper, our
expectation is that this geometry should be some kind of $11$ dimensional LLM geometry which is analog of $10d$ LLM geometry. This $11d$ LLM
geometry should arise from the some deformation of $AdS_4 \times S^7$ geometry. Since we have new geometry, it is expected that  $1/N$
expansion should be reorganized like ${\cal N}=4$ SYM theory. We found that the new expansion is $1/(N+M)$ where $M$ is the number of columns
in the representative Young diagram of the operator and its order of $N$.\\

\noindent
{\bf Acknowledgements:} We would like to thank Robert de Mello Koch for pleasant discussions at every step of the work, going through the draft and making valuable comments on the initial draft to improve the quality of the final draft.

\end{document}